\documentclass[revtex,prb,twocolumn,superscriptaddress,showpacs]{revtex4-1}
\usepackage{amsmath}   
\usepackage{amssymb}
\usepackage{braket}
\usepackage{leftidx}
\usepackage{graphicx}    
\usepackage{verbatim}   
\usepackage{xcolor}         
\usepackage{subfigure}  
\usepackage{hyperref}   
\usepackage{dcolumn}    
\usepackage{textcomp}
\usepackage{siunitx}
\usepackage{threeparttable}
\usepackage{todonotes}
\usepackage{titlecaps}
\usepackage{multirow}
\usepackage{lipsum}
\begin{document}


\title{Multiscale simulations of uni-polar hole transport in (In,Ga)N quantum well systems}

\author{Michael~O'Donovan}
\email{michael.odonovan@tyndall.ie}
\affiliation{Tyndall National Institute, University College Cork, Cork T12 R5CP, Ireland}\affiliation{Department of Physics, University College Cork, Cork T12 YN60, Ireland}
\author{Patricio~Farrell}
\affiliation{Weierstrass Institute (WIAS), Mohrenstr. 39, 10117 Berlin, Germany}
\author{Timo~Streckenbach}
\affiliation{Weierstrass Institute (WIAS), Mohrenstr. 39, 10117 Berlin, Germany}
\author{Thomas~Koprucki}
\affiliation{Weierstrass Institute (WIAS), Mohrenstr. 39, 10117 Berlin, Germany}
\author{Stefan~Schulz}
\affiliation{Tyndall National Institute, University College Cork, Cork T12 R5CP, Ireland}

\date{\today}

\begin{abstract}
Understanding the impact of the alloy micro-structure on carrier transport becomes important when designing III-nitride-based LED structures. In this 
work, we study the impact of alloy fluctuations on the hole carrier transport in (In,Ga)N single and multi-quantum well systems. To disentangle hole 
transport from electron transport and carrier recombination processes, we focus our attention on uni-polar ($p$-$i$-$p$) systems. The calculations employ
our recently established multi-scale simulation framework that connects atomistic tight-binding theory with a macroscale drift-diffusion model. In 
addition to alloy fluctuations, we pay special attention to the impact of quantum corrections on hole transport.
Our calculations indicate that results from a virtual crystal approximation present an upper limit for the hole transport in a $p$-$i$-$p$ structure in terms of the current-voltage characteristics. Thus we find that alloy fluctuations can have a detrimental effect on hole transport in (In,Ga)N quantum well systems, in contrast 
to uni-polar electron transport. However, our studies also reveal that the magnitude by which the random alloy results deviate from virtual crystal approximation data depends on several factors, e.g. how quantum corrections are treated in the transport calculations.
\end{abstract}

\maketitle


\section{Introduction}

The semiconductor alloy indium gallium nitride ((In,Ga)N) has attracted significant research interest for optoelectronic device applications due to its 
in principle flexible band gap engineering across the visible spectral range.~\cite{Hump2008}
In general, (In,Ga)N alloys have several unique features which are not found in other III-V material systems (e.g.\ (In,Ga)As). 
Firstly, heterostructures such as (In,Ga)N/GaN quantum wells (QWs) grown along the wurtzite $c$-axis exhibit strong internal electrostatic
built-in fields across the QW.~\cite{AmMa2002,MiCa2013} Such fields are absent in (In,Ga)As/GaAs wells grown along the [001]-direction of their underlying
zincblende structures. The built-in field in $c$-plane (In,Ga)N/GaN QWs is induced by spontaneous polarization, as 
well as a strain related piezoelectric contribution.~\cite{AmMa2002,MiCa2013} A consequence of this internal electric field is (i) a decrease in electron-hole 
wavefunction overlap and (ii) a red-shift in the emission wavelength; this is also known as the quantum confined Stark effect.~\cite{WiSc2009}
Secondly, and  equally important for this study,
(In,Ga)N alloys and connected heterostructures display strong carrier localization effects even for a random alloy 
micro-structure.~\cite{WaGo2011,ScCa2015,TaDa2020} This effect is particularly strong for holes, which have a higher effective mass than the
electrons.\cite{ScCa2015}

As such, understanding the impact of the alloy fluctuations on carrier transport becomes important when designing (In,Ga)N-based 
LED structures. In order to gain insight into the connection between alloy fluctuations and carrier (electron and hole) transport in
(In,Ga)N-based multi-quantum well (MQW) systems, studying the properties of uni-polar structures present a very promising and interesting alternative to investigating a full
LED structure. Previous works have focused already on uni-polar electron transport in $n$-doped-intrinsic-$n$-doped ($n$-$i$-$n$) (In,Ga)N/GaN MQW
systems.\cite{BrMa2015,MiOD2021_JAP} These investigations revealed that alloy fluctuations\cite{BrMa2015,MiOD2021_JAP} as well as quantum effects~\cite{MiOD2021_JAP} are
important for describing the electron transport, leading for instance to a lower knee/turn-on voltage of the device and an improved theory experiment comparison for such systems.
However, far less attention has been directed towards uni-polar \emph{hole} transport.~\cite{ShWe2021} This stems in part from the fact that high
quality $p$-doped-intrinsic-$p$-doped ($p$-$i$-$p$) systems are challenging to realise experimentally (high dopant activation energy\cite{PeKo2000_JAP}, 
compensation effect\cite{DaIi2010_JCG}, memory effect \cite{YO_JCG_1994}), but also from 
the fact that the theoretical modelling of carrier localization in (In,Ga)N systems is a difficult task in itself.~\cite{DiVPe2020,ChODo2021} 

Here, we apply our previously established multi-scale simulation framework,~\cite{MiOD2021_JAP} that bridges the gap between atomistic electronic structure theory and macroscale drift-diffusion (DD) carrier transport simulations, to study uni-polar hole transport in (In,Ga)N single QW (SQW) and MQW systems. We analyze in detail the impact of alloy and quantum corrections on the results. Our calculations reveal that in contrast to previously reported uni-polar \emph{electron} transport results, alloy fluctuations have a detrimental effect on the hole transport in (In,Ga)N MQWs. 

The manuscript is organized as follows: In Sec.~\ref{sec:Theory} we outline the theoretical framework
we use, introducing the underlying tight-binding (TB) model as well as localization landscape theory (LLT) and the DD
settings. In Sec.~\ref{sec:Results} we present our results for uni-polar hole transport in (In,Ga)N/GaN SQW and MQW
systems. Finally, Sec.~\ref{sec:Conclusions} concludes this work.

\section{Theoretical framework}\label{sec:Theory}
In this section we outline briefly the main ingredients of our theoretical framework. A detailed discussion is given 
in Ref.~\onlinecite{MiOD2021_JAP}. In Sec.~\ref{subsec:TB} we introduce
the TB model and a ``local'' TB Hamiltonian that is used to obtain the local band edge energies. We then 
describe briefly in Sec.~\ref{subsec:mesh_generation} how the band edge energy is transferred and connected to the device
simulation mesh used in the DD solver. The DD model underlying the calculations is presented in
Sec.~\ref{subsec:DDmodel}. 

\subsection{Tight-binding model and energy landscape generation}
\label{subsec:TB}

The theoretical framework starts with a $sp^3$ nearest-neighbour TB model which is 
described in detail in Refs.~\onlinecite{MiCa2013}
and~\onlinecite{ScCa2015}. This approach, combined with valence force field and local polarization models, allows us to 
capture the impact of (random) alloy fluctuations on the electronic structure of (In,Ga)N QW systems on an atomistic 
scale. While it is possible to use such an atomistic electronic structure theory as the backbone for carrier transport calculations,\cite{ODoLu2021,JGeSPr_PSS_2018} it is computationally very expensive to simulate a full device structure. To reduce the computational load, while still keeping 
essential atomistic information, we proceed as follows. In a first step, we 
extract an energy landscape from TB which can be used in the active 
region of a device. Since we are here interested in uni-polar hole transport, our active region consists of  
(In,Ga)N QWs; in a full LED structure, the active region may also include an (Al,Ga)N electron blocking layer. Outside 
the active region a coarser mesh resolution, as described in more detail below, is used. This is motivated by 
the fact that the barrier regions are made up of binary GaN, which does not exhibit alloy 
fluctuations. In order to extract an energy landscape, we construct a ``local'' TB Hamiltonian 
from the full TB Hamiltonian, and diagonalize it at each lattice site of the simulation cell that describes the active region.  
In doing so one can extract local valence (or conduction) band edge energies which contain (local) strain and polarization effects arising
from alloy fluctuations. More details on the method are given in Ref.~\onlinecite{ChODo2021}. Equipped with such an energy landscape either 
electronic structure or carrier transport calculations can be performed using modified continuum-based models.~\cite{ChODo2021,MiOD2021_JAP} 

\subsection{Device mesh generation, smoothing alloy fluctuations and quantum corrections}
\label{subsec:mesh_generation}

In this subsection, we discuss in detail key aspects of our approach to connect the TB energy landscape to drift-diffusion simulations. First, we describe the device mesh structure. At its core lie two different types of meshes: an atomistic and a significantly coarser macroscopic mesh. The former corresponds to the QW/active region which the latter embeds into a device.
In the following, we detail different types of smoothing operations on the atomistic mesh. We smooth the atomistic valence 
band edge (VBE) data obtained from TB either via Gaussian averaging, LLT or a combination of both operations. Gaussian averaging and LLT, help to account for quantum effects which classical DD theory does not directly consider.
In the final subsection, we pay particular attention to subtleties of applying LLT in a MQW case. 



\subsubsection{Device mesh structure} 
\label{sec:device_mesh_structure}

Our device mesh consists of an atomistic and macroscopic part.
The atomistic mesh corresponds to the single or multi-well quantum region. Since we will solve the LLT equation on this mesh via finite element method (FEM), see below, we refer to it as FEM mesh as well. In this mesh, each node location and data site correspond to the position of an atom and its VBE energy, respectively. Since our goal is to study macroscopic DD currents, we embed the atomistic mesh into a macroscale device mesh with doped contact regions on either side. Our goal has two immediate implications. Since the doped regions are a couple of orders of magnitude larger than the QW region and do not exhibit alloy fluctuations, the mesh in these regions can be chosen to be significantly coarser; this helps to reduce the computational cost. Moreover, DD simulations are typically performed via the finite volume method (FVM). Here, in particular, we use the Voronoi FVM.\cite{Farrell2017} Since this method requires a boundary-conforming tetrahedral Delaunay mesh, we not only enlarge the QW mesh by introducing meshes for the doped regions but also insert a few additional points within the QW region itself. Atomistic VBE data is then interpolated onto these additional nodes. Within the doped regions on either side of the QW region, we set uniform (GaN) VBE data. All atomistic nodes within the FEM mesh are also included in the FVM mesh. Both meshes are created via \texttt{TetGen}\cite{Si15ACM} and the interpolation is handled via \texttt{WIAS-pdelib}.\cite{pdelib} The device mesh generation is explained visually and in more detail in Ref. \onlinecite{MiOD2021_JAP}.

\subsubsection{Smoothing by Gaussian averaging}

Previously it has been discussed~\cite{LiPi2017} that the spatial scale over which alloy fluctuations are relevant for carrier transport is 
linked to the de Broglie wavelength of the carriers. Given the semi-classical and continuum-based nature of ``standard''
DD models, such effects are not captured. To remedy this shortcoming, in a first step we employ a Gaussian averaging procedure on the FEM mesh given by

\begin{equation}
    E_{v}^{\sigma}(\mathbf{x_i}) = \frac{\sum_j E_{v}^\text{TB}(\mathbf{x_j})
    \times \exp\Big(\frac{-|\mathbf{x_i}-\mathbf{x_j}|^2}{2\sigma}\Big)}{\sum_j
    \exp\Big(\frac{-|\mathbf{x_i}-\mathbf{x_j}|^2}{2\sigma}\Big)}\,\, .
    \label{eq:Gaussian_Broadening}
\end{equation}

The averaging procedure accounts now for the effect that carrier wavefunctions do not only ``see'' valence or 
conduction band energies at a given lattice site but also beyond this. In doing so, the averaging procedure 
depends now on the width of the Gaussian, $\sigma$. We note that the above is similar to Ref.~\onlinecite{LiPi2017},
however our approach differs in that we average here the TB band edge energy, $E_{v}^\text{TB}$,
which contains already local strain and built-in field effects obtained on an atomistic level; 
in Ref.~\onlinecite{LiPi2017} first local  In, Ga contents are determined and then, using continuum elasticity theory, 
local strain and built-in potentials are evaluated before the local band edge energy values are calculated. 
Given that the Gaussian width $\sigma$ is now effectively a free parameter, we will study below the impact of $\sigma$ on the 
effective energy landscape and the hole transport. Future studies may target
evaluating $\sigma$ values based on calculations of e.g.\ the density of states~\cite{PiLi2017,McTa2020} in (In,Ga)N-based QWs utilizing modified 
continuum models. 

\begin{figure}[t!]
    \centering
    \includegraphics[width=0.5\textwidth]{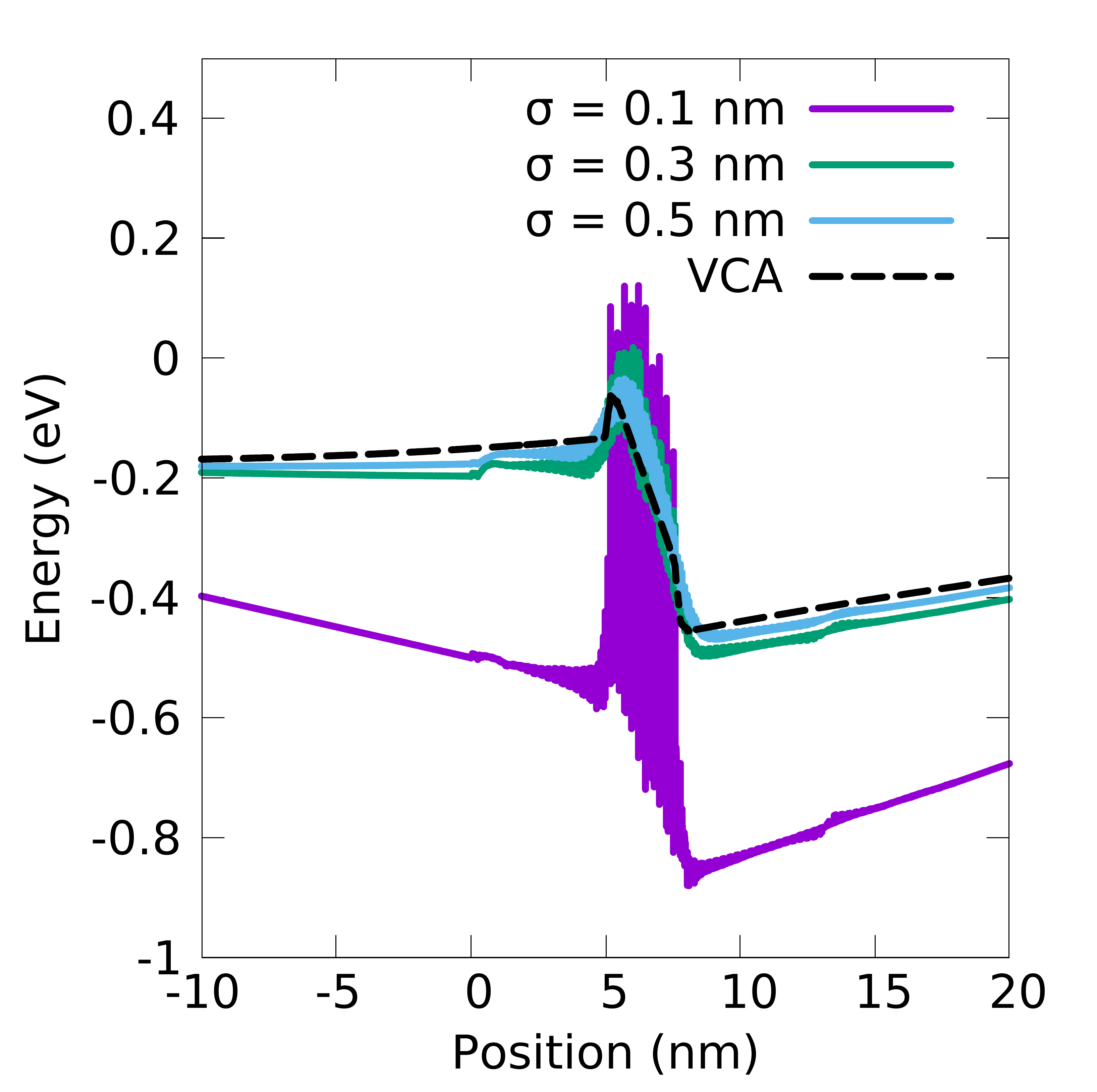}
    \caption{Comparison of valence band edge energies for an In$_{0.1}$Ga$_{0.9}$N single quantum well of width 3.1~nm at a bias of 0~V (equilibrium solution) without 
    quantum corrections for a VCA (black, dashed) and random alloy calculations
    using a Gaussian width, $\sigma$, of 0.1~nm (purple), 0.3~nm (green) and 0.5~nm (blue).
    }
    \label{fig:VBE_SQW_NoLLT}
\end{figure}

To understand the potential impact of $\sigma$ on the results, Fig.~\ref{fig:VBE_SQW_NoLLT} shows the VBE energy profile of an (In,Ga)N/GaN SQW with 10\% In and a width of 3.1~nm for different 
values of $\sigma$ ($\sigma = 0.1$~nm (purple), $\sigma = 0.3$~nm (green) and $\sigma = 0.5$~nm (blue)) at
equilibrium (0V). The VCA profile, which does not undergo broadening, is also depicted (black, dashed). 
Firstly, we note that when choosing a $\sigma$ value smaller than the bond length of the material, $d_0$
(e.g. $\sigma = 0.1$~nm $< d_0^{GaN}$),\cite{DSPTanner_thesis}
basically no averaging takes place.
As a consequence, the VBE energy exhibits very strong fluctuations due to 
the alloy fluctuations, see Fig.~\ref{fig:VBE_SQW_NoLLT}. We note that while the average QW ``depth'' (averaged
over each atomic plane) does not differ significantly for different $\sigma$ values, both the magnitude of the 
VBE energy fluctuations as well as the potential barrier between (In,Ga)N well and surrounding GaN is strongly 
impacted by the $\sigma$ value. 
Thus, Fig.~\ref{fig:VBE_SQW_NoLLT} gives already indications that carrier transport, e.g.\ current voltage (I-V) curves, may be 
strongly dependant on $\sigma$. We will discuss this in more detail below.

\subsubsection{Quantum corrections by localization landscape theory}

While the above introduced Gaussian averaging procedure provides local corrections to the confining energy landscape, 
it does not provide information about the electron and hole ground state energy in a QW system since it is not coupled with a quantum mechanical description by e.g.\ solving the Schr\"odinger equation. On the other hand, 
most conventional/commercial transport simulators often have the option to couple DD simulations with solving Schr\"odinger's equation, however, they neglect alloy fluctuations.
As discussed for instance in detail in Ref.~\onlinecite{LiPi2017}, it is numerically very demanding to studying carrier transport in (In,Ga)N/GaN LED structures when treating alloy fluctuations and quantum corrections in a fully 3-D self-consistent Schr\"odinger-Poisson-DD framework. To this end we employ the numerically far more efficient localization
landscape theory (LLT)\cite{ArDa2016,FiPi2017,ChKe2020} to account for quantum corrections in our 3-D simulations. Thus instead of solving Schr\"odinger's equation, we solve the LLT equation supplied with Dirichlet and Neumann boundary conditions:
\begin{equation}
    \hat{H}^\text{EMA}u = \left(-\frac{\hbar^2}{2m^\star}\Delta + (V - E_\text{ref})\right)u = 1\,\, .
    \label{eq:LLT}
\end{equation}
Here $m^\star$ denotes the effective mass, $\hbar$ is Planck's constant, and $E_\text{ref}$ is the reference 
energy of the system. The choice of $E_\text{ref}$ will be discussed in detail in Sec.~\ref{subsec:problems_with_LLT_in_MQW}.
$V$ is the confining potential which is extracted from the local band edge energy 
values: since we are targeting uni-polar hole transport, $V$ is determined by the VBE energy. 
As the derivation of LLT requires that $\hat{H^\text{EMA}}$ is a positive definite operator,\cite{MaFi2012_NAS}
Eq. (\ref{eq:LLT}) is solved in the hole picture (where the hole ground state has the \emph{lowest} energy on an absolute scale, and 
the hole effective mass is \emph{positive}) rather than in the valence band picture (where the hole ground state has 
the \emph{highest} valence band energy on an absolute scale, and the hole effective mass is \emph{negative}). As such, the confining
potential is described by $V = -E^\text{TB}_v$.


We note that LLT involves solving a linear partial differential equation instead of a large eigenvalue problem as in case of the Schr\"odinger equation. Therefore, LLT facilitates a numerically more efficient 3-D carrier transport simulation framework. We solve the LLT equation numerically with \texttt{WIAS-pdelib},\cite{pdelib} with more details given in Ref.~\onlinecite{MiOD2021_JAP}.

To include quantum corrections via LLT into our transport calculations, we make use of the fact that once $u$ is determined from solving Eq.~(\ref{eq:LLT}), an effective potential, $W$, which describes the 
localization landscape of the confining potential $V$
can be extracted at each mesh-point via:\cite{MaFi2012_NAS,FiPi2017}
\begin{equation}
    W(\mathbf{x}_i) = \frac{1}{u(\mathbf{x}_i)} + E_\text{ref}\,\, .
    \label{eq:W=1/u}
\end{equation}
As the effective landscape is obtained in the hole picture, it is converted to the valence band picture (multiplication by -1) so that it can be used in transport calculations. \emph{When displaying band edge profiles, we always use the valence band picture; if LLT has been applied, the obtained effective landscapes/potentials, $W$, have been transformed accordingly.}


\begin{figure}[t!]
    \centering
    \includegraphics[width=0.5\textwidth]{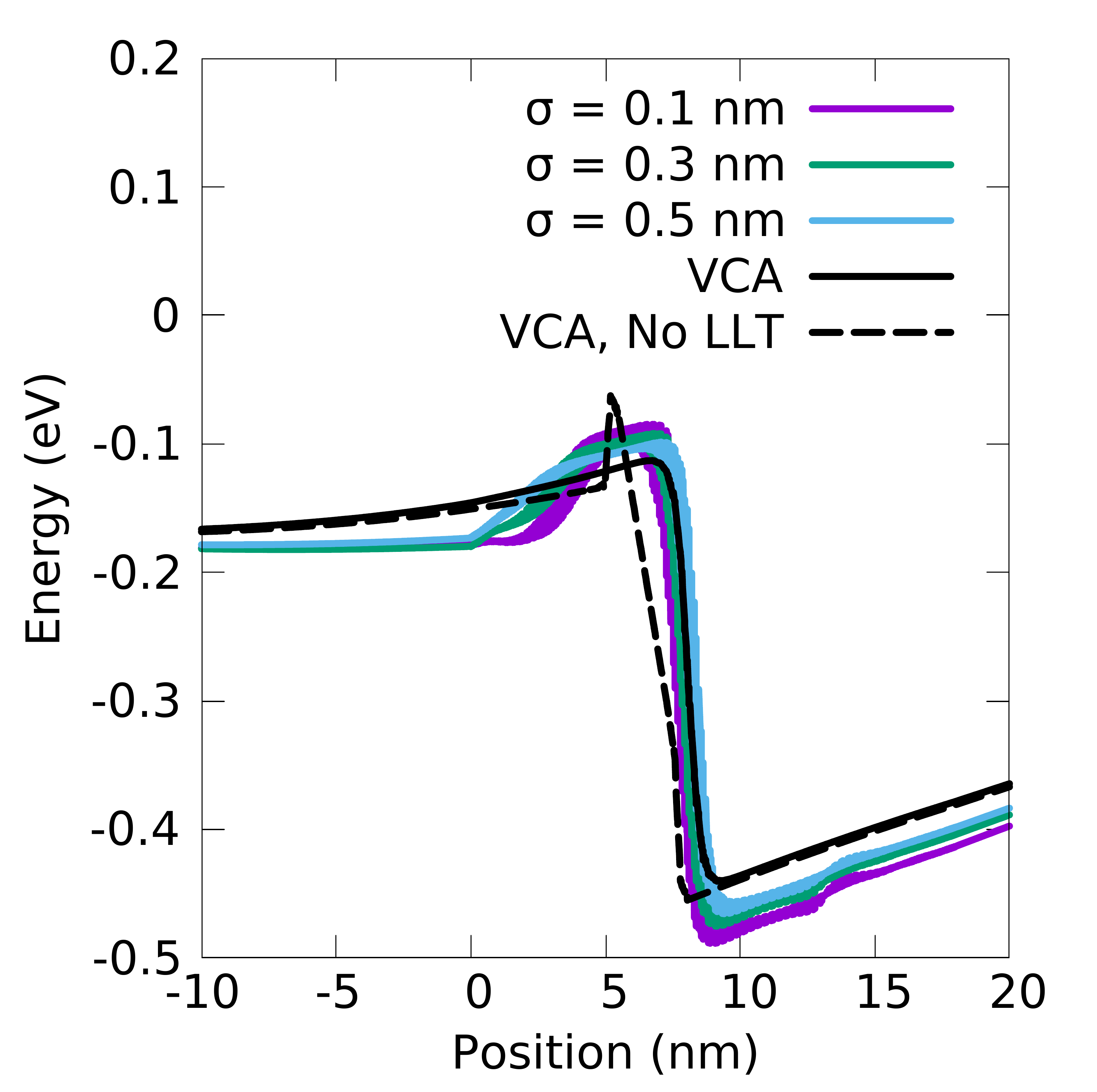}
    \caption{Comparison of valence band edge energies for a In$_{0.1}$Ga$_{0.9}$N single quantum well of width 3.1~nm at 0~V \emph{including}
    quantum corrections via LLT for a VCA (black, solid) and random alloy calculations
    using a Gaussian width of 0.1~nm (purple), 0.3~nm (green) and 0.5~nm (blue).
    The VCA result excluding quantum corrections is also shown (black, dashed).}
    \label{fig:VBE_SQW_LLT}
\end{figure}


To provide a first general insight into the impact of LLT corrections to the confining energy landscape for carriers, 
Fig.~\ref{fig:VBE_SQW_LLT} shows the effective potential $W$ for the VBE of an In$_{0.1}$Ga$_{0.9}$N/GaN SQW system at
equilibrium (0V); the width of the well is 3.1~nm.
The data are displayed for three different Gaussian broadening values $\sigma$, namely $\sigma=0.1$~nm (purple), $\sigma=0.3$~nm (green) and $\sigma=0.5$~nm (blue), as well as a LLT corrected VCA profile (black, solid). A ``standard'' VCA profile is also shown (black, dashed). This figure displays that once LLT is included in the calculations the impact of $\sigma$ on the band edge profile is significantly reduced. Looking at the VCA plus LLT results, one finds a very smooth confining band edge energy profile.
The consequences of using a softened profile for carrier transport will be discussed below.

\subsubsection{Subtleties of LLT for MQW structures}\label{subsec:problems_with_LLT_in_MQW}

Before turning to our DD framework and how we use $W$ in it, we discuss first some subtleties of the LLT approach, which become important when dealing with (In,Ga)N \emph{MQW} systems. To calculate $W$, one has to solve Eq.~(\ref{eq:LLT}) to obtain $u$ first.
As discussed in Refs.~\onlinecite{FiPi2017} and~\onlinecite{ChKe2020}, $u$ can be written as an 
expansion of the eigenstates $\psi_j(\mathbf{x_i})$ of the system under consideration:
\begin{equation}
    u(\mathbf{x_i}) = \sum_j \alpha_j \psi_j(\mathbf{x_i})\,\, .
    \label{eq:u_expansion}
\end{equation}
The expansion coefficients $\alpha_j$ are then given by
\begin{equation}
    \alpha_j = \sum_{\mathbf{x_i}\in \Omega} \frac{\psi_j(\mathbf{x_i})}{E_j}\,\, .
    \label{eq:expansion_coeffs}
\end{equation}
From Eqs.~(\ref{eq:u_expansion}) and~(\ref{eq:expansion_coeffs}) it is apparent that $u$ and thus the resulting 
effective potential $W$ depends on the magnitude of the energy eigenvalues $E_j$ of a given $\psi_j(\mathbf{x_i})$ and its energetic separation to other (higher lying) states. Thus, if for instance the ground state energy $E_0$ is small (close to 0) and the energy separation to higher lying states $E_j$ with $j\neq 0$ is large, $u$ describes basically the ground state wavefunction (and ground state energy) as one can see form Eq.~(\ref{eq:u_expansion}). As a consequence, $u$ is a very good approximation of lowest energy state in a given ``localization'' region $\Omega$. On the other hand, if $E_0$ is large and energetically close to higher lying states, $u$ may contain contributions not only from the ground state but also higher lying states. To achieve, on an absolute scale, a small ground state energy one may adjust the energy scale of the system by choosing an appropriate \emph{reference} energy $E_\text{ref}$ such that $E_0-E_\text{ref}  > 0 $ is small  compared to the energy separation with higher lying states. In doing so $u$ and thus the effective potential $W$ is dominated by the ground state wavefunction of e.g.\ an (In,Ga)N QW.

\begin{figure}
    \centering
    \includegraphics[width=0.5\textwidth]{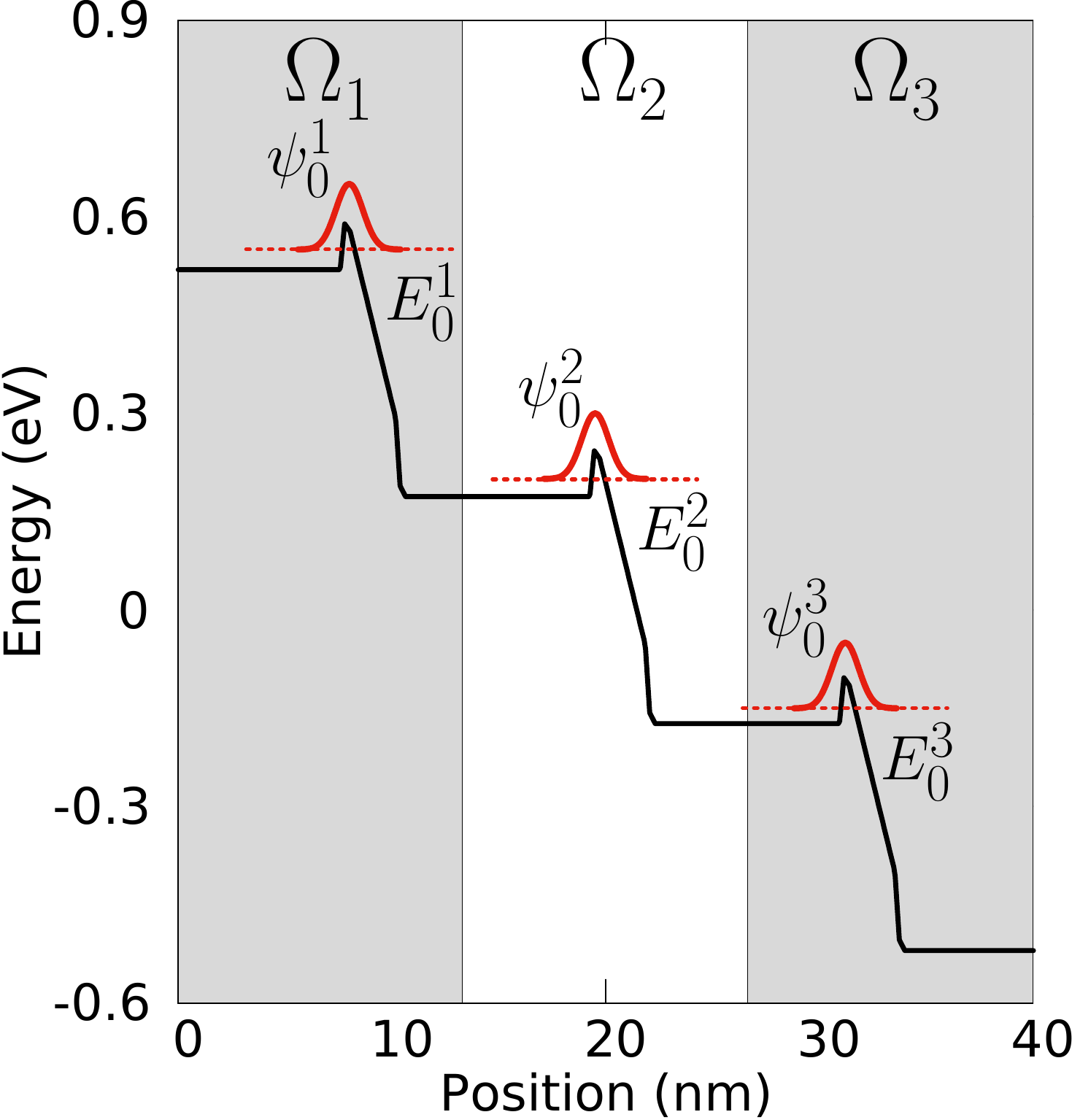}
    \caption{Schematic illustration of a potential band edge energy profile (black solid line) in a multi-quantum well with 3 quantum wells
    where the 
    wells exhibit a large energy separation between their respective ground state energies $E^i_0$ (red dashed line). The local 
    hole ground state wavefunction in the i$^{th}$ localization region, $\Omega_i$ (marked by shading), are indicated by 
    $\psi^i_0$ (red, solid).} 
    \label{fig:cartoon_psi}
\end{figure}

While the above can be realised in a straightforward manner for a $SQW$ systems (simply using the lowest (highest) CBE (VBE) energy as $E_\text{ref}$), for a MQW system this becomes more involved.
To illustrate this in more detail, Fig.~\ref{fig:cartoon_psi} shows a schematic of a 3 QW system. Here we assume a large energy difference between the VBE values of the different wells to highlight central aspects of LLT. If this structure is treated as one single ``localization'' region $\Omega$, and we choose the reference energy, $E_\text{ref}$, to be very close to $E_0^1$ (using the hole picture instead of the valence picture), $u$ and consequently $W$ will be dominated by the ground state wavefunction $\psi^1_0$, as $\alpha_0\psi_0$ will dominate the series expansion in Eq.~(\ref{eq:u_expansion}), originating from Eq.~(\ref{eq:expansion_coeffs}). Due to the larger energy separation between $E_\text{ref}$ and $E_0^2$ and $E_0^3$, respectively, there will be basically no contribution from $\psi^2_0$ and $\psi^3_0$ to $W(\mathbf{x_i})$. As a consequence the effective potential $W(\mathbf{x_i})$ in the spatial region where $\psi^2_0$ (located in region $\Omega_2$) and $\psi^3_0$ (located in region $\Omega_3$) are localized is largely unaffected by LLT quantum corrections. 

To circumvent this issues, one could in principle partition the system into multiple (here three) subregions ($\Omega_1,\Omega_2, \Omega_3$) and solve LLT for each sub-system separately; for each subregion an individual $E_\text{ref}^i$ can be chosen. In doing so, the wavefunctions $\psi_0^i$ describe now the ground state wavefunction for each ``localization'' region $\Omega_i$ with its corresponding local ground state energy $E_0^i$. Now the series expansion of $u$ in each region 
is dominated by the first term, and $u$ obtained for each region $\Omega_i$ should give a very good description of the lowest 
state locally. As a consequence, the confining potential in each QW subregion $\Omega_i$ contains quantum corrections.

When using this approach of partitioning the system into different subregions, the remaining question is how to ``connect'' the local effective potentials $W_i$ so that one obtains a global one, $W$. In the case of electrons, partitioning the system 
into different localization regions is difficult, as the low effective electron 
mass leads to a large ``leaking'' of the wavefunction into the barrier material. This makes it very difficult to connect the individual effective potentials. Further discussions on consequences of the effective confinement potential for \textit{electron} transport can be found in Ref.~\onlinecite{MiOD2021_JAP}.
Holes, however, have a much higher effective mass, and partitioning the system is achievable if the separation between the wells in a MQW system is not too small. For the system under consideration (see Sec.~\ref{sec:Results}) this is the case and the locally obtained effective landscapes return quickly to the band edge energy of the GaN barrier material; this guarantees that the interface 
between neighbouring localization regions is smooth and continuous when ``stitching'' the different $W_i$ together to obtain $W$. 
A comparison of effective landscapes obtained with and without partitioning a MQW structure into different sub-regions is show in Appendix~\ref{Appendix:LLT_partitioning_landscapes}.
When analyzing hole carrier transport in a MQW system in Sec.~\ref{sec:MQW}, we will pay special attention to the above described partitioning of the system when including quantum corrections via LLT in the simulations.

\subsection{Uni-polar drift-diffusion model}
\label{subsec:DDmodel}

As discussed in Section \ref{sec:device_mesh_structure}, we transfer the atomistic VBE data, together with constant macroscopic VBE parameters for the doped regions, on to a FVM mesh. Following the discussion in the previous section, we may use for the atomistic VBE data either $E_v^{\sigma}(\mathbf{x}_i)$, see Eq.~(\ref{eq:Gaussian_Broadening}), or 
$-W(\mathbf{x}_i)$, 
see Eq.~(\ref{eq:W=1/u}); the multiplication of $W$ by $-1$ is due to the change from the hole picture to the valence band picture.
Next, we present the DD models which describe charge transport through our device. 


Charge carrier transport is modelled using the van Roosbroeck system.\cite{VanRoosbroeck1950} As we are interested in uni-polar hole transport, the stationary van Roosbroeck system consists of two coupled nonlinear partial differential equations of the form:

\begin{subequations}
\label{eq:vanRoosbroeck}
\begin{align}
 -\nabla \cdot \left(\varepsilon_s(\mathbf{x}) \nabla \psi(\mathbf{x})\right) &=q\left( p(\mathbf{x}) + C\right)\,\, ,
 \label{eq:poisson-equation} \\
 \nabla\cdot \mathbf{j}_p &= 0 
\end{align}
\label{eq:vRS}
\end{subequations}
for $\mathbf{x} \in \Omega$. The Poisson equation, Eq.~(\ref{eq:poisson-equation}), describes the electric field generated
by the scalar electric potential $\psi(\mathbf{x})$ in the presence of a free (hole) charge carrier density, $p(\mathbf{x})$. 
Here, $\varepsilon_s(\mathbf{x}) = \varepsilon_0\varepsilon_r(\mathbf{x})$ describes the position dependent dielectric
constant; $q$ is the elementary charge. In a (doped) uni-polar semiconductor device, the overall charge density is given by the
density of free (positively charged) holes, $p(\mathbf{x)}$, and the density of ionized built-in dopants, 
$N_A^+(\mathbf{x})$, where $C = -N_A^+(\mathbf{x})$ denotes the density of singly ionized acceptor atoms. The current 
density $\mathbf{j}_p(\mathbf{x})$ is given by~\cite{Farrell2017}

\begin{equation}
    \mathbf{j}_p(\mathbf{x}) = -q\mu_p p(\mathbf{x}) \nabla\varphi_p(\mathbf{x})\,\, .
    \label{eq:hole_current}
\end{equation}

That is, the negative gradient of the quasi Fermi potential, $\varphi_p(\mathbf{x})$, is the driving force of the 
current; $\mu_p(\mathbf{x})$ denotes the free carrier (hole) mobility.

Using the Boltzmann approximation, the densities of free carriers, $p(\mathbf{x})$, in a solid are given by

\begin{align}
p(\mathbf{x}) &= N_v \exp\left(
\frac{q(\varphi_p(\mathbf{x}) - \psi(\mathbf{x})) + E_v^{dd}(\mathbf{x})}{k_B T} \right),
\label{eq:carrier-densities}
\end{align}

where $k_B$ is the Boltzmann constant, $T$ denotes the temperature, $E_v^{dd}(\mathbf{x})$ is the (position dependent)
VBE energy and $N_v$ is the effective density of states:

$$N_v = 2\Bigg(\frac{2\pi m^*_h k_B T}{\hbar^2}\Bigg)^{3/2}.$$

The VBE energy $E_v^{dd}$ in the DD simulations
may now be chosen to be (smoothed) TB data, $E_v^{dd} = E_v^{\sigma}$, VCA data, $E_v^{dd} = E_v^\text{VCA}$, or the outcome of the LLT calculations, $E_v^{dd} = -W$. In doing so the VBE energy $E_v^{dd}$ may vary spatially due to random alloy fluctuations. Thus, care must be taken to discretize the hole flux correctly. To this end we extend the well-known Scharfetter-Gummel flux approximation\cite{Scharfetter1969} to 
variable band edge energy values, as detailed in Ref.~\onlinecite{MiOD2021_JAP}.
Bias values are implemented via Dirichlet boundary conditions. Details of this 
approach can be found in Ref.~\onlinecite{Farrell2017}.


\section{Results}\label{sec:Results}

\begin{table}[t]
    \centering
    \begin{tabular}{|l|l|c|}
    \hline
    Physical Quantity & Value & Units \\
    \hline
    $m_h^\star$ GaN & 1.87  & m$_0$\\
    $m_h^\star$ InN     & 1.61 & m$_0$\\
    $\mu_h\ p-$GaN & 5 & cm$^2$/(V s) \\
    $\mu_h\ i-$GaN & 10$^\dagger$ & cm$^2$/(V s) \\
    $\mu_h\ i-$(In,Ga)N & 10 & cm$^2$/(V s) \\
    $\epsilon_r^\text{GaN}$ & 9.7 & $\epsilon_0$ \\
    $\epsilon_r^\text{InN}$ & 15.3 & $\epsilon_0$ \\
    $p-$doping (GaN) & 2$\times$10$^{19}$ & cm$^{-3}$ \\
    \hline
    \end{tabular}
    \caption{Material parameters used in the simulations. Unless otherwise specified, all parameters are taken from Ref.~\onlinecite{LiPi2017};
    $^\dagger$ Ref.~\onlinecite{CKLi_IEEE_2014}.}
    \label{tab:material_parameters}
\end{table}

\begin{figure}[t]
    \centering
    \includegraphics[width=0.5\textwidth]{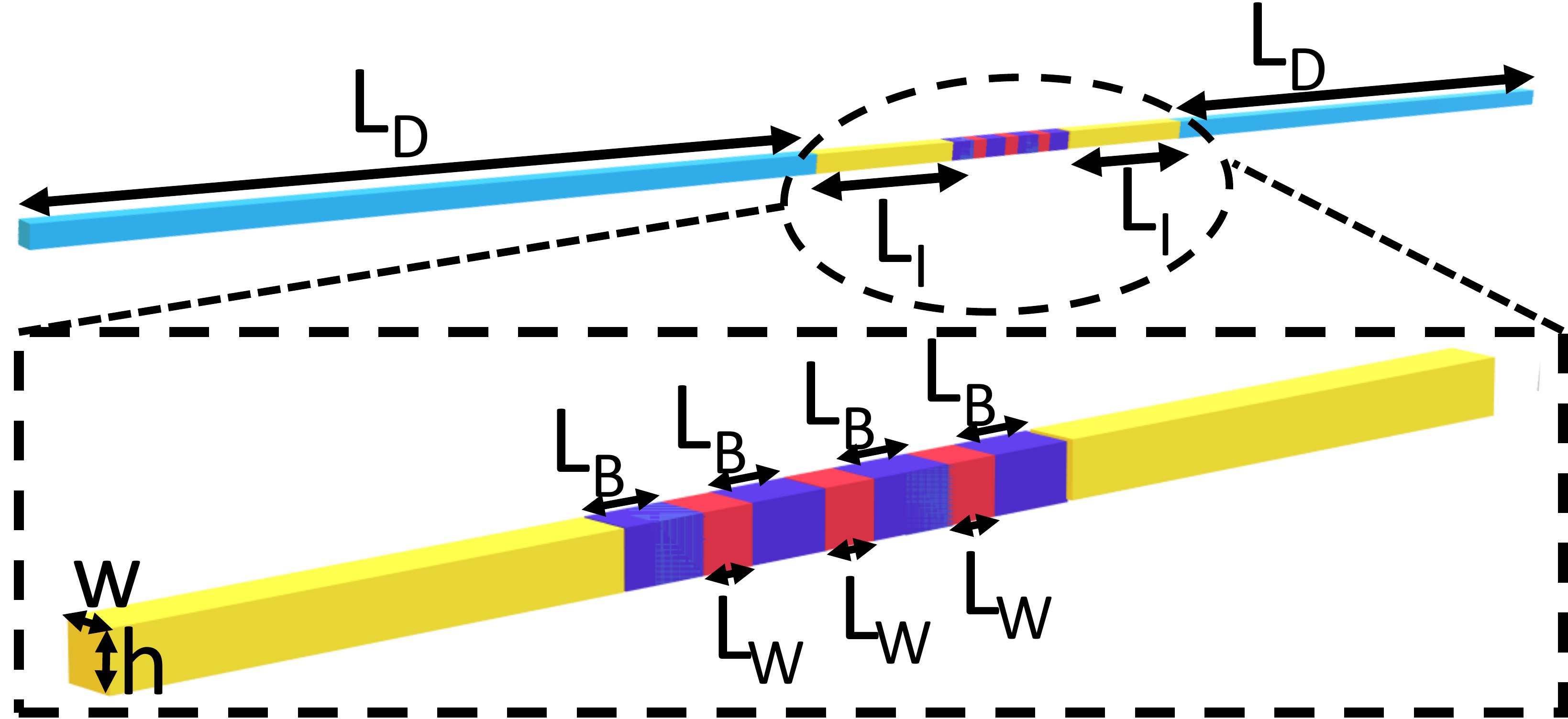}
     \caption{Schematic illustration of the simulation cell with three quantum wells
(QWs) in the active region. The $p$-doped regions (light blue) have a doping
density of $2\times10^{19}$ cm$^{-3}$ and a length of $L_D =$ 160~nm. The intrinsic
regions on the coarse mesh (yellow) have a length of $L_I = 40$~nm. The atomistic
region, also assumed as intrinsic, contains regions of a GaN barrier material
(dark blue) with a length of $L_B =8.0$~nm and In$_{0.1}$Ga$_{0.9}$N QWs (red) with a
width of $L_W =3.1$~nm. For a single QW calculation the atomistic region contains only one In$_{0.1}$Ga$_{0.9}$N QW ($L_w=3.1$~nm) and two GaN barrier regions. The simulation cell has an in-plane dimension of $w \times h =5.1 \times 4.4$~nm$^2$ along the entire system. 
}
    \label{fig:schematic}
\end{figure}

In this section, we apply the framework described above to a $p$-doped-intrinsic-$p$-doped ($p$-$i$-$p$) system in both a SQW, Sec.~\ref{sec:SQW}, and a MQW, Sec.~\ref{sec:MQW}, setting. Our simulations are carried out within the \texttt{ddfermi} simulation tool\cite{ddfermi} which is implemented within the \texttt{WIAS-pdelib} toolbox.\cite{pdelib} A schematic of the MQW system including the contact regions
is shown in Fig.~\ref{fig:schematic}. Details about well and barrier widths, as well as the In content are given in the figure caption. The material parameters entering the DD calculations are summarized in Table~\ref{tab:material_parameters}; all calculations have been performed at a temperature of $T=300$ K. To study the influence of alloy fluctuations and quantum corrections on the carrier transport, the simulations have been performed for the different $E_v^{dd}$ settings discussed in Sec.~\ref{subsec:DDmodel}. Thus, we compare results from calculations including alloy fluctuations to results from VCA simulations; the simulations have been carried out in the absence and presence of LLT quantum corrections. In the case of the MQW, we also investigate how the current-voltage (I-V) curves change when partitioning the MQW system to solve LLT locally (for each QW).

\subsection{Single QW}\label{sec:SQW}

\begin{figure}[t]
    \centering
    \includegraphics[width=0.5\textwidth]{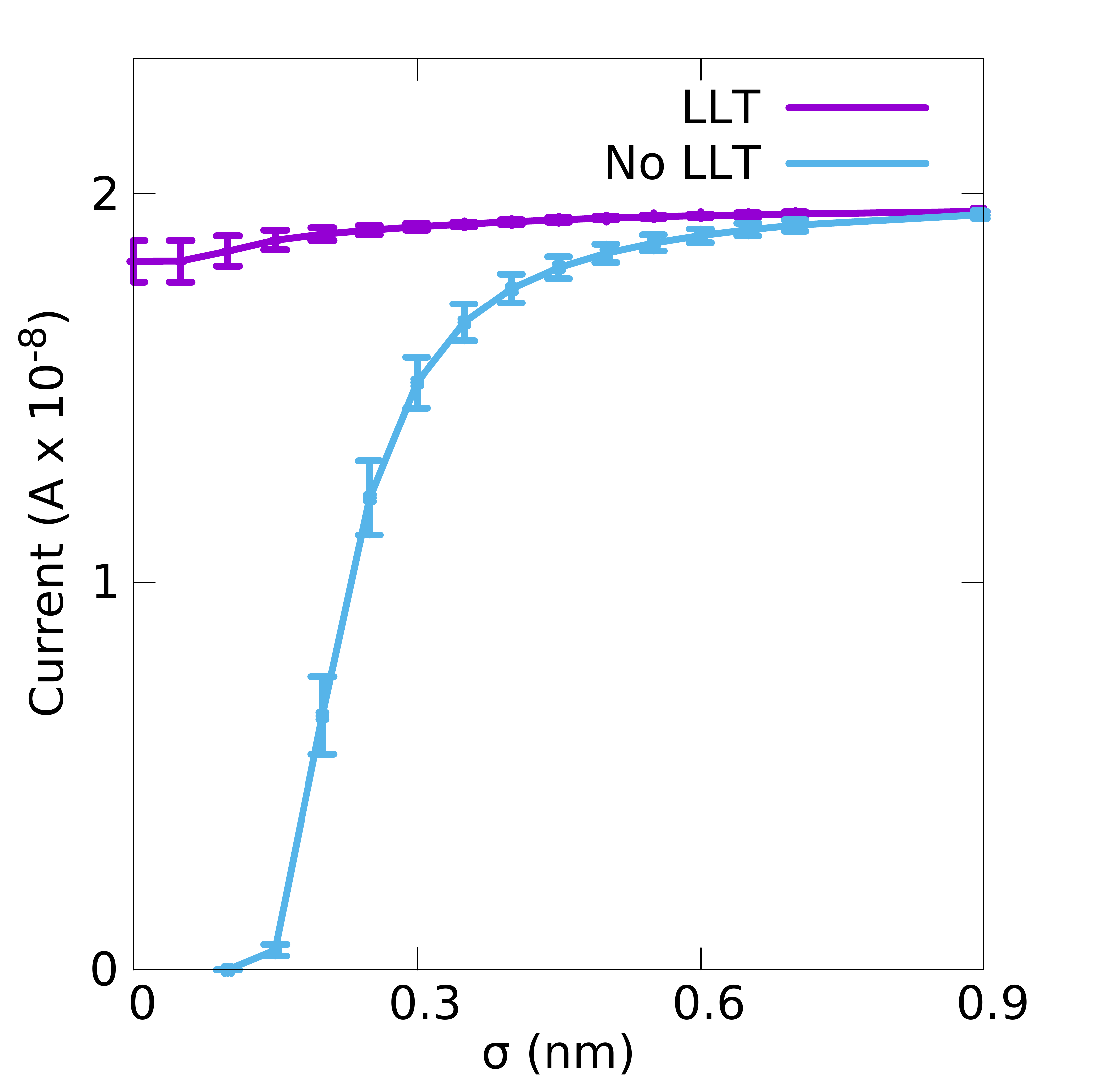}
    \caption{Impact of Gaussian width, $\sigma$, on the current in a single In$_{0.1}$Ga$_{0.9}$N/GaN quantum well system at a bias of 3.0~V. Results are obtained in the presence (purple) and absence (blue) of quantum corrections via LLT and 
    are averaged over 5 different microscopic configurations. The errorbars
    show the standard deviation of the current over the 5 configurations. }
    \label{fig:sigma_vs_J_SQW}
\end{figure}

In the following we analyze the impact of random alloy fluctuations and quantum corrections on the I-V
characteristics of a $p$-$i$-$p$ (In,Ga)N SQW system; details of the structure and simulation cell are given in the caption of Fig.~\ref{fig:schematic}.
In order to study the influence of the
alloy microstructure on the results we have repeated these calculations for 5 different microscopic configurations. 
Furthermore, the Gaussian broadening $\sigma$ has been varied to study how $\sigma$ affects the 
results. Before turning our attention to
the full I-V curve of the system, Fig.~\ref{fig:sigma_vs_J_SQW} depicts the current in the SQW
system at a fixed bias of 3V for different $\sigma$ values. As discussed in Sec.~\ref{subsec:mesh_generation}, when $\sigma$ is increased, the Gaussian function 
softens the VBE and reduces the magnitude of the fluctuations. As consequence, in the \emph{absence} of quantum corrections, the 
current at 3V increases with increasing $\sigma$ and starts to converge for $\sigma$ values larger than approximately 0.5~nm. 
For these large $\sigma$ values the VBE becomes smooth and the current approaches that of a completely smooth VCA landscape (not shown).
In addition, Fig.~\ref{fig:sigma_vs_J_SQW} also reveals that there is an abrupt increase in the current at around $\sigma=0.2$~nm. We attribute 
this to the fact that if $\sigma$ is small 
and below the bond length of e.g.\ GaN, the band edge profile entering the DD simulations exhibits
strong (local) fluctuations which noticeably affect the carrier transport.  

\begin{figure}[t]
    \centering
    \includegraphics[width=0.5\textwidth]{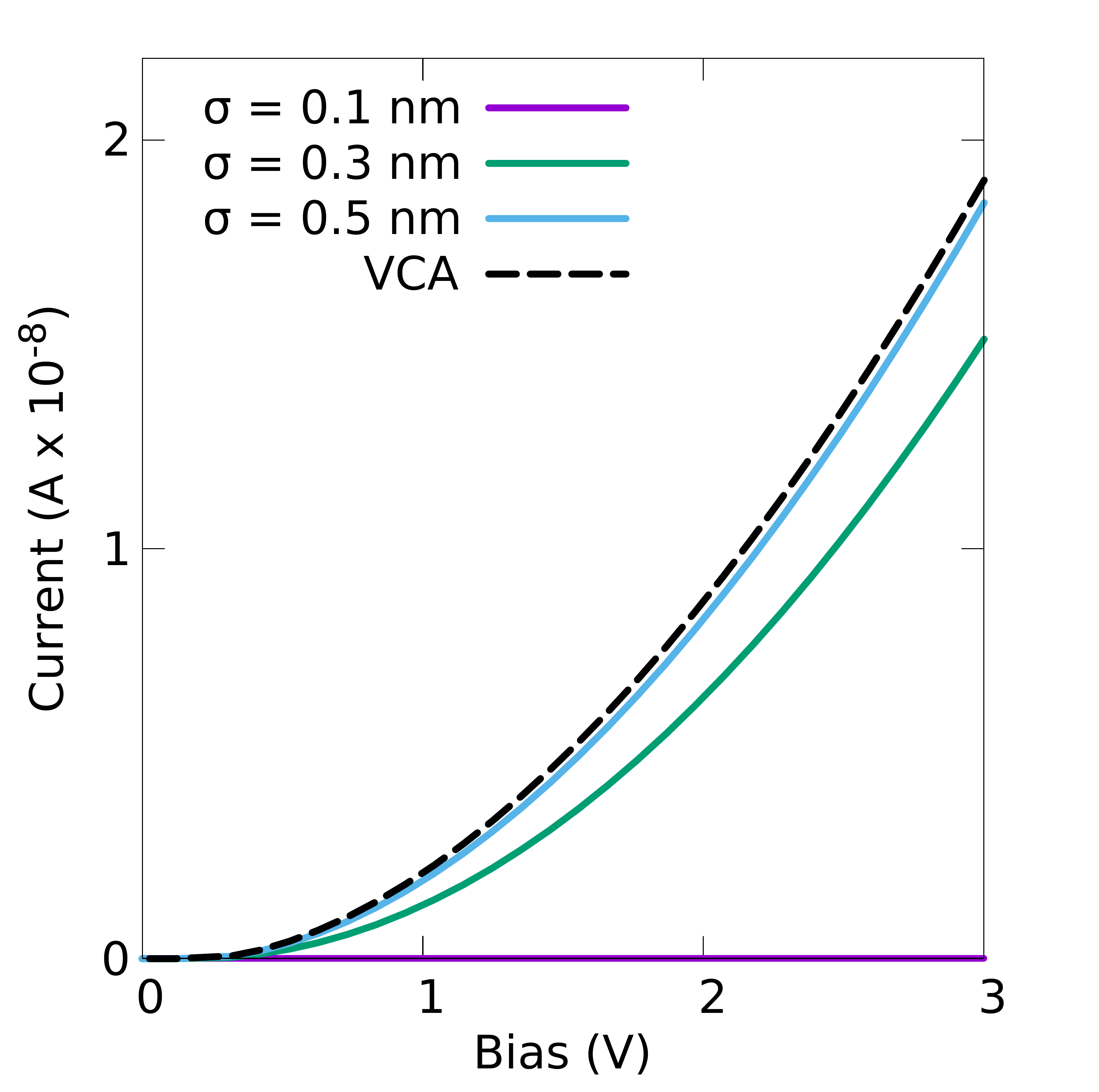}
    \caption{Comparison of current-voltage curves for a single In$_{0.1}$Ga$_{0.9}$N/GaN quantum well 
    for VCA (black, dashed) and random alloy calculations
    using a Gaussian width of $\sigma=0.1$~nm (purple), $\sigma=0.3$~nm (green) and $\sigma=0.5$~nm (blue)
    in the absence of quantum corrections.
    }
    \label{fig:IVs_SQW_NoLLT}
\end{figure}

In the next step we turn our attention to the full I-V curves in the presence of alloy fluctuations but the 
\emph{absence} of LLT quantum corrections. Overall, the behavior discussed for the fixed bias of 3V, Fig.~\ref{fig:sigma_vs_J_SQW}, 
is also reflected in the full I-V curves, Fig.~\ref{fig:IVs_SQW_NoLLT}: for a Gaussian width of $\sigma = 0.1$~nm the current is
extremely low, but increases with increasing $\sigma$. However, it is important to note that the here obtained results are in contrast to uni-polar \emph{electron} transport, for instance discussed in Ref.~\onlinecite{MiOD2021_JAP}. In the case of the electrons, the current always \emph{exceeds} the VCA results, while we find here that in the hole case it \emph{approaches} the VCA data. This means that for electron transport alloy fluctuations are beneficial, while they are detrimental for the hole transport. This result is consistent with the observation that alloy fluctuations lead to strong hole localization effects, while electron wavefunctions, due to their lower effective mass, are affected to a lesser extend by the alloy fluctuations.\cite{WaGo2011,ScCa2015}

\begin{figure}[t]
    \centering
    \includegraphics[width=0.5\textwidth]{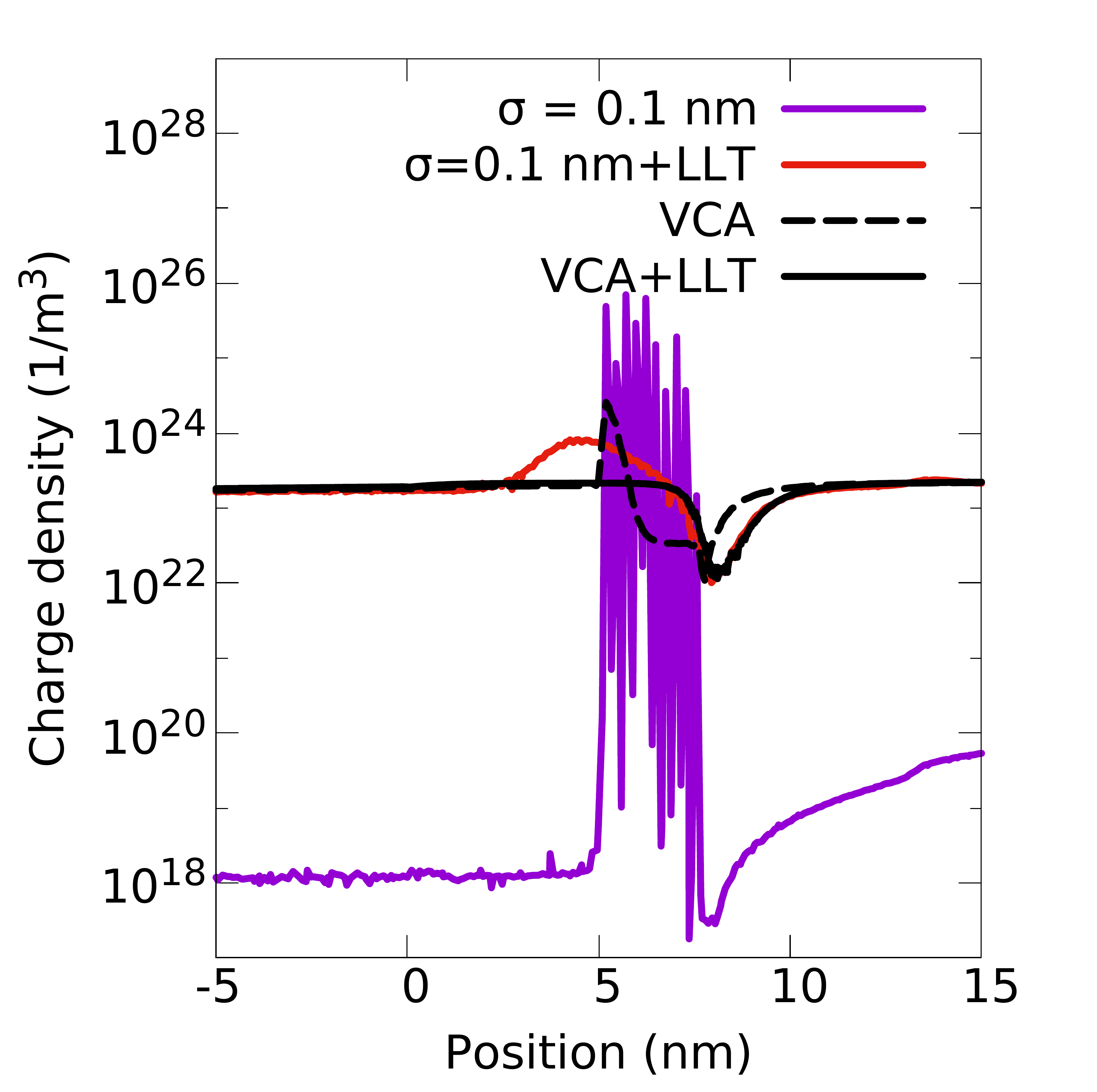}
    \caption{Carrier density distribution in and around a single In$_{0.1}$Ga$_{0.9}$N/GaN quantum well of width 3.1~nm at a bias of 2.9~V
    for calculations including random alloy fluctuations and using a Gaussian width of 0.1~nm. The results are shown in the absence (purple) and presence (red) of quantum corrections via LLT. For comparison VCA data (black, dashed), and VCA 
    including LLT (black, solid) are also depicted. }
    \label{fig:rho_SQW_NoLLT_0p1}
\end{figure}

To shed more light onto the influence of alloy fluctuations on the hole transport, Fig.~\ref{fig:rho_SQW_NoLLT_0p1} shows the charge density distribution in and around the (In,Ga)N SQW region for $\sigma = 0.1$~nm in the absence of any LLT quantum corrections and at a bias of 2.9~V. For comparison the VCA charge density distribution is also depicted (black, dashed) and the VCA charge density distribution including quantum corrections (black, solid). We stress again that due to the small $\sigma$ value, the alloy fluctuations lead to a strongly fluctuating VBE energy profile, which in turn results in strong hole localization effects.
From Fig.~\ref{fig:rho_SQW_NoLLT_0p1} one can infer that due to the strong carrier localization effect, we observe a very high carrier density, 
when compared to the VCA result, in the QW region; the carrier density in the barrier material is depleted in the random alloy case compared to VCA. 
As a consequence, these carrier localization effects/the strong VBE fluctuations lead to a strong VBE bending, originating from the coupling
of the hole density 
and the quasi-Fermi level via Eq.~(\ref{eq:hole_current}) and Eq. (\ref{eq:carrier-densities}). Overall, and compared to the VCA result, this gives rise to a larger resistivity of the device. Thus for this small value of $\sigma=0.1$~nm, the current through the device is very low, as seen in Fig.~\ref{fig:IVs_SQW_NoLLT}.
We note that such a low broadening parameter can result in an underlying energy landscape which
is not compatible with the DD framework (as $\sigma$ is much lower than the de Broglie wavelength), and this extreme depletion of the barriers may
be physically unrealistic.



\begin{figure}[t]
    \centering
    \includegraphics[width=0.5\textwidth]{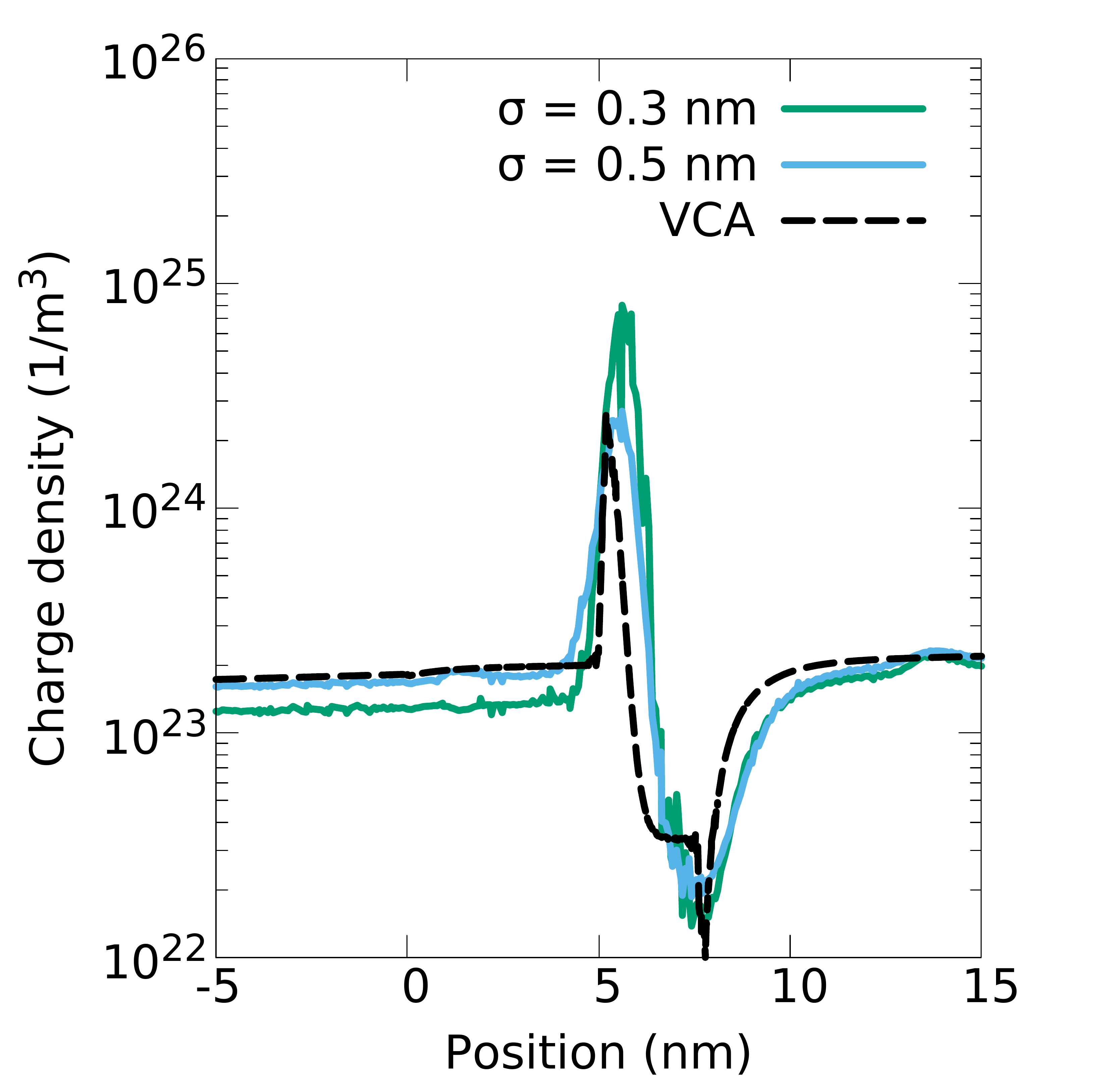}
    \caption{Carrier density distribution in and around a single In$_{0.1}$Ga$_{0.9}$N/GaN quantum well of width 3.1~nm at a bias of 2.9~V
    for a VCA (black, dashed) and random alloy calculations.
    The latter use Gaussian widths of $\sigma=0.3$ nm~(green) and $\sigma=0.5$~nm (blue) and exclude
    quantum corrections. }
    \label{fig:rho_SQW_NoLLT_no0p1}
\end{figure}

The situation changes with increasing $\sigma$ as Fig.~\ref{fig:rho_SQW_NoLLT_no0p1} shows. Here, the charge density distribution in and around the QW for both $\sigma = 0.3$~nm (green) and $\sigma = 0.5$~nm (blue) 
are similar to the VCA results (black, dashed). Furthermore, as the charge density distributions with 
increasing $\sigma$ approaches the VCA profile, so does the resulting I-V curve,
Fig.~\ref{fig:IVs_SQW_NoLLT}.



\begin{figure}[htbp]
    \centering
    \includegraphics[width=0.5\textwidth]{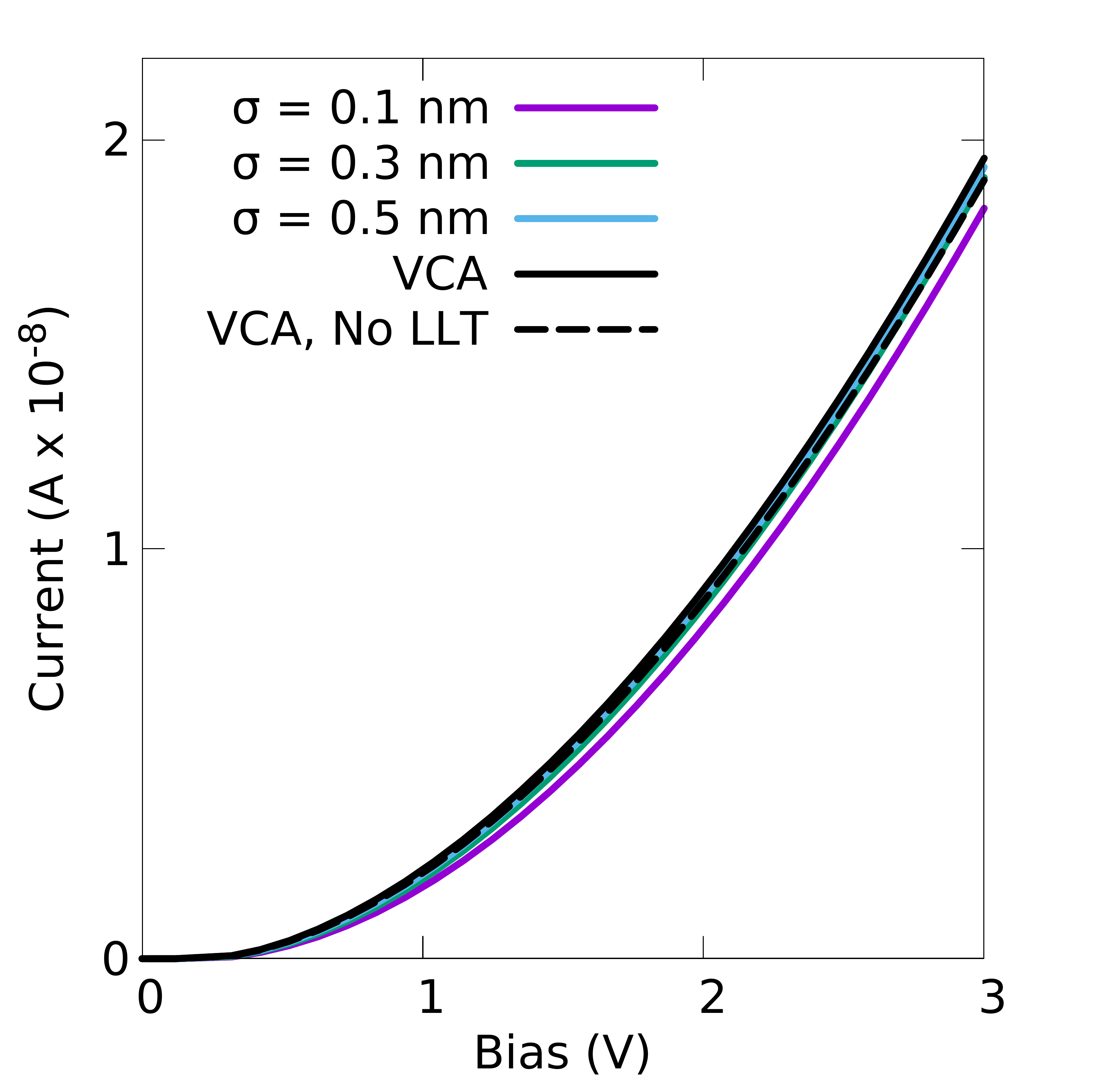}
    \caption{Including quantum corrections via LLT: Comparison of current-voltage curves for a single In$_{0.1}$Ga$_{0.9}$N/GaN quantum well
    of width 3.1~nm for a VCA (black, solid) and random alloy calculations; the random alloy simulations
    use Gaussian widths of $\sigma=0.1$~nm (purple), $\sigma=0.3$~nm (green) and $\sigma=0.5$~nm (blue).
    }
    \label{fig:IVs_SQW_LLT}
\end{figure}


Having discussed the impact of alloy fluctuations on the hole transport, we focus our attention now on the impact of quantum 
corrections on the results. Overall, we find that when including quantum corrections via LLT in the transport calculations, 
the Gaussian width $\sigma$ influences the results to a much lesser extent.
This can for instance been seen in Fig.~\ref{fig:sigma_vs_J_SQW}, where the current is shown as a function of $\sigma$ (purple line) at a fixed bias of 3 V. 
In contrast to the 
results without quantum corrections (light blue), when including these 
corrections, the obtained current changes very little when increasing $\sigma$ beyond 0.2~nm. We highlight also that even at the very low $\sigma$ value of $\sigma=0.1$~nm, the current is strongly increased when including quantum corrections. The origin of this becomes clear when looking again at the carrier density profile in and around the SQW, depicted in Fig.~\ref{fig:rho_SQW_NoLLT_0p1}. As discussed above, in the absence of quantum corrections, the strongly fluctuating energy landscape leads to a very large carrier density in the well and depletes the region surrounding the well. When accounting for quantum corrections, the carrier density profile including alloy fluctuations (red), even though the same $\sigma$ value is used, is much smoother and approaches the VCA quickly in the barrier. This emphasizes again that quantum corrections soften the confining energy landscape and indicates that once LLT corrections are taken into account, the alloy microstructure is of secondary importance for the carrier transport. This is confirmed by 
Fig.~\ref{fig:sigma_vs_J_SQW}: the standard deviation (indicated by the error bars in the figure) is small relative to the current, at least for larger $\sigma$.  The impact of the  alloy microstructure is still  visible for smaller $\sigma$ values. We note here also that the magnitude of this effect may depend on the in-plane dimension of the simulation cell, especially when using small $\sigma$ values. Thus careful studies are required to analyze this in more detail, including a further evaluation on the choice of the ``correct'' Gaussian width before  LLT is applied. 

When turning to the full I-V curve of the SQW system, Fig.~\ref{fig:IVs_SQW_LLT}, we find that the choice of $\sigma$ is of secondary importance, at least for the studied system. In addition to the random alloy calculations, Fig.~\ref{fig:IVs_SQW_LLT} depicts also
VCA results both in the presence (black, solid) and absence (black, dashed) of LLT quantum corrections.
From this it is clear that in the case of a SQW, random alloy results do not differ strongly from the VCA data. Interestingly, these results are also well approximated by VCA simulations \emph{excluding} quantum corrections. 
For the VCA, when there are no alloy fluctuations and the VBE is smooth, the combination of the small valence band offset as well
as the high hole effective mass, results in similar profiles for the confining potentials of the VCA and quantum corrected VCA. Consequently
the I-V curves do not differ significantly.

It should be noted that the above discussed results are different but also similar to uni-polar \emph{electron} transport. They are similar in the sense that once quantum corrections are taken into account, VCA and random alloy simulations give very similar results in terms of the I-V characteristics of SQW systems. However, a difference between electron and hole transport is that for uni-polar electron transport the current increases with increasing $\sigma$ and exceeds the VCA result, for holes this is not the case. Our calculations also indicate that for holes, once LLT corrections are included, the current is basically independent of $\sigma$. However, it should again be noted that this result may depend on the in-plane dimensions of the simulation cell. A larger in-plane cell may give rise to a larger extent of locally varying band edge energies. As a consequence carrier localization effects may be more pronounced. Thus the here presented results should be treated as ``best'' case scenario, since when carriers are ``trapped'' by alloy fluctuations they will increase the resitivity of the device. We conclude therefore that in general carrier localization effects will have a detrimental effect on the hole transport, and the resulting currents will in general be smaller or equal to the VCA result, in contrast to electrons.



However, the impact of carrier localization effects on the I-V curves may be more pronounced in MQWs, as the depletion of the carriers in 
the GaN barrier region may be amplified. In our previous study on uni-polar \emph{electron} transport we have already seen that results 
from a SQW system cannot necessarily be 
carried over to MQWs. 
In general, gaining insight into hole transport in MQW systems is very important for understanding the carrier distribution in full (In,Ga)N-based 
MQW LED structures. Thus, we turn our attention in the next section to uni-polar hole transport in (In,Ga)N MQW structures.

\subsection{Multi QW}\label{sec:MQW}

\begin{figure}[t]
    \centering
    \includegraphics[width=0.5\textwidth]{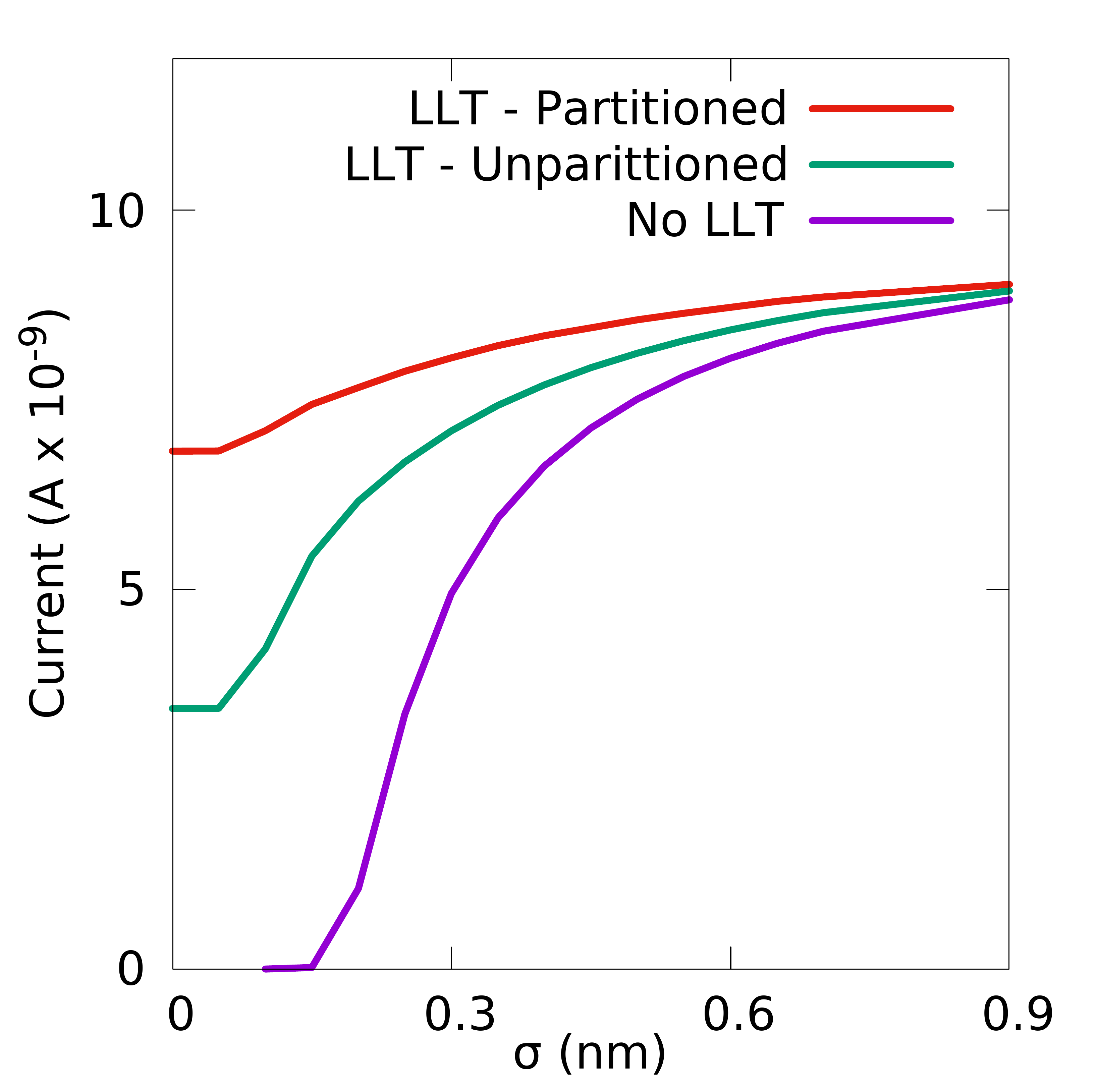}
    \caption{Impact of Gaussian width, $\sigma$, on the current in an In$_{0.1}$Ga$_{0.9}$N/GaN multi-quantum well system at 3.0~V. Results are
    shown for a system including quantum corrections via LLT and partitioning the system
    into 3 localization regions each with a local reference energy (red), for a system including 
    quantum corrections via LLT using a single (global) reference energy for the entire multi-quantum well region (green),
    and for a system excluding quantum corrections (purple).
    }
    \label{fig:sigma_vs_J_MQW}
\end{figure}

Similar to the SQW system discussed in the previous section, we start our analysis of the hole transport in a (In,Ga)N/GaN 
MQW system by investigating the impact of the Gaussian width $\sigma$ on the results. 
Figure~\ref{fig:sigma_vs_J_MQW} displays the current through the MQW system as a function of $\sigma$ at a fixed bias of 3V. 
Here we compare results from simulations that (i) exclude quantum corrections via LLT (purple), (ii) include quantum corrections via LLT but treating the entire MQW region as
one localization region (green), and (iii) quantum corrections via LLT but solving the LLT equation for each well of the MQW system
separately (red), as discussed in Sec.~\ref{subsec:mesh_generation} (cf. Fig.~\ref{fig:cartoon_psi}).


Figure~\ref{fig:sigma_vs_J_MQW} shows that for all studied $\sigma$ values, the calculation excluding LLT (purple line) exhibits the lowest 
current at a fixed voltage of 3V. Also, the difference is largest at small $\sigma$ values. In the case of the calculation without 
LLT corrections the VBE edge exhibits large local fluctuations. These fluctuations are intrinsically smoothed by the quantum corrections, and 
the resulting landscape (even for small $\sigma$ values) exhibits significantly smaller fluctuations due to the alloy microstructure. 
The large VBE fluctuations increase the potential barrier and consequently increase the resistance in the 
$p$-$i$-$p$ junction thus leading to a smaller current. This is the same effect we have already seen in the SQW system, however 
the result is more pronounced due to the combined influence of the 3 QWs in the MQW. 


In a second step we discuss the results from the calculations including quantum corrections in more detail. Looking at the 
simulations using a global reference energy, i.e. the MQW system is treated as a single localization region (green),
we find that the current drops a greater amount at low $\sigma$ values compared the the outcome of the simulations using a local reference 
energy (here each well is treated as a separate localization region). More specifically, at the smallest considered $\sigma$ 
value (no broadening), the current 
obtained from the model using a global reference energy is just over half the current using local reference energies. We 
attribute this drop to the combination of two factors. Firstly, given that the LLT model using a local reference energy also 
shows a slight drop in current with decreasing $\sigma$ indicates that the strong fluctuations in the VBE energy still impact the current even though the LLT treatment softens this intrinsically.
Secondly, when treating the MQW as a single localization region the confining potential of the QW for which the VBE energy 
is furthest away from the global reference energy is expected to be poorly described in such an LLT treatment. As a consequence, still larger fluctuations are present in the wells furthest away from the reference energy, especially for small $\sigma$ values. All this will 
result in a higher resistivity of the MQW system and consequently a lower current at fixed bias. 


\begin{figure}[t]
    \centering
    \includegraphics[width=0.5\textwidth]{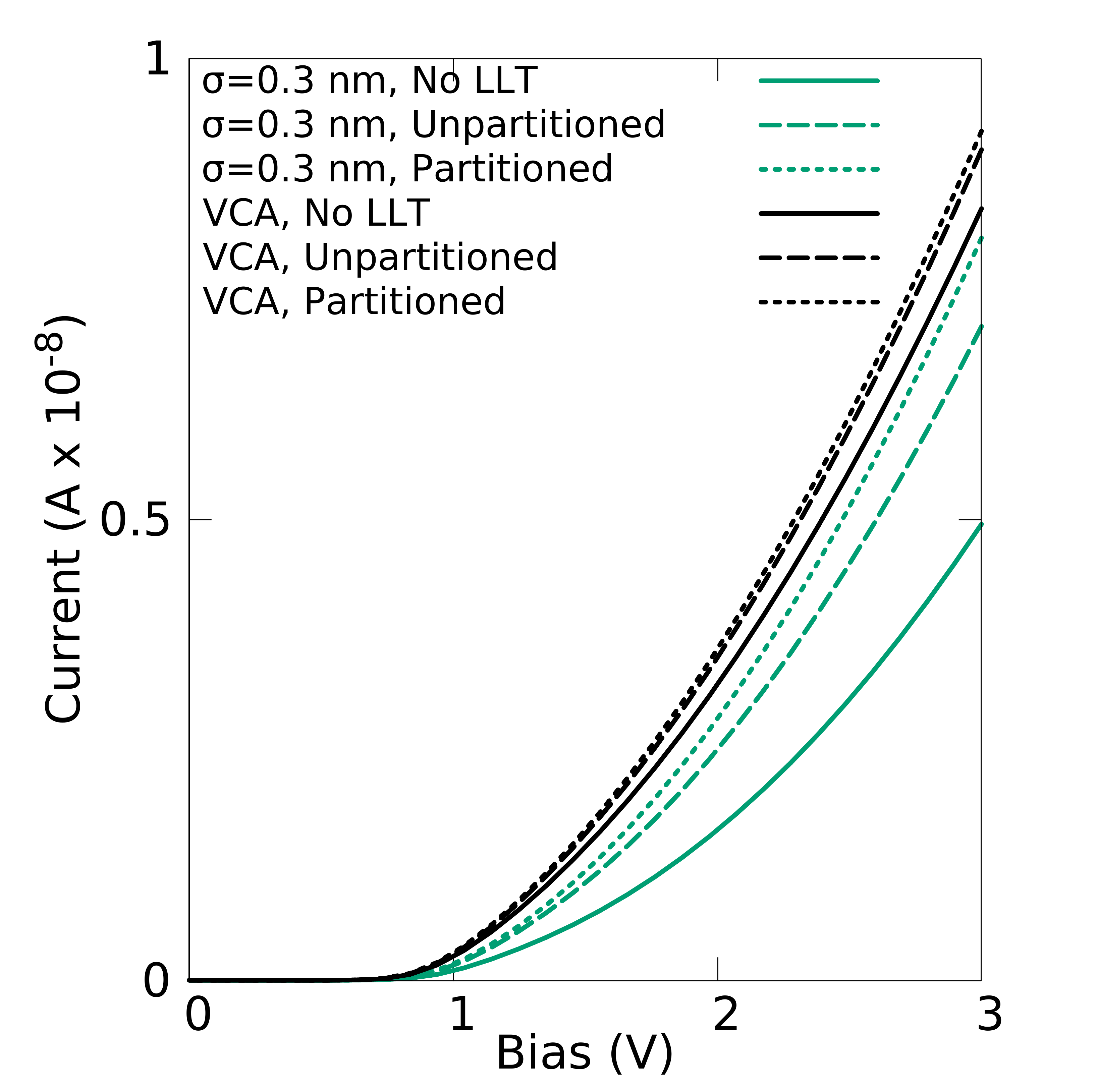}
    \caption{Comparison of current-voltage curves in a multi-quantum well In$_{0.1}$Ga$_{0.9}$N/GaN system for VCA (black)
    and random alloy calculations; the random alloy simulations use a Gaussian width of $0.3$~nm (green). I-V curves are 
    shown for calculations without any quantum corrections (solid), including quantum 
    corrections when employing an un-partitioned (dashed) and  partitioned multi-quantum well regions (dotted).
    }
    \label{fig:IVs_MQW}
\end{figure}

Having discussed the impact of Gaussian broadening and LLT quantum corrections on the current in a MQW system at a 
fixed bias, Fig.~\ref{fig:IVs_MQW} depicts the full I-V curves. Here again results from calculations applying LLT, 
both using a single localization region (dashed), $\Omega$, and sub-regions, $\Omega_i$, for each QW (dotted), as well as results in the 
\emph{absence} of quantum corrections (solid) are shown. This is displayed for 
both VCA (black) and random alloy calculations using a Gaussian width of $0.3$~nm (green); to get first insight into the hole transport in a MQW structure we have restricted the calculations 
to one alloy configuration. Future studies can target analysing the statistics of
different alloy microstructure configurations on the results. A value of 
$\sigma = 0.3$~nm has been chosen since it is large enough for the Gaussian averaging to including neighbouring sites but
small enough to still capture effects due to carrier localization. 
Figure~\ref{fig:IVs_MQW} reveals that in both VCA and random alloy calculations, quantum corrections increase the current similar to the situation in uni-polar electron transport.~\cite{MiOD2021_JAP} 
Furthermore, when using a local reference energy for LLT, thus treating each QW as an individual localization region, $\Omega_i$, the 
current increases further when compared to the LLT model using a global reference energy. Our results also show that this effect is more pronounced for the random alloy case; partitioning the system in VCA impacts the I-V curve (black dashed and black dotted line) only slightly. 

Overall our calculations reveal that in the MQW system and for the chosen $\sigma$ value of $\sigma=0.3$~nm, even when including LLT corrections, the random alloy calculations give a smaller current at fixed bias when compared to the VCA result. This finding is in contrast to the SQW system, where VCA and random alloy results give very similar results, see Fig.~\ref{fig:IVs_SQW_LLT}. Furthermore, and again in contrast to the SQW structure, the magnitude of the difference in current between VCA and random alloy results will depend on the $\sigma$ value, as Fig.~\ref{fig:sigma_vs_J_MQW} shows. Future studies targeting for instance theory experiment comparisons are now required to gain further insight into the broadening parameter $\sigma$. We note that beyond $\sigma$, and as already mentioned bove, the in-plane dimension of the simulation cell may impact the results as
carrier localization effects due to lateral fluctuations in the alloy can have a (detrimental) influence on the current.
Furthermore, it should be noted that the LLT treatment builds on a single-band effective mass approximation; our previous studies indicate that such a model may underestimate hole localization effects,~\cite{ChODo2021} which in turn may lead to higher current.

Nevertheless, all these factors should only reduce the current further in the MQW system. Thus the VCA I-V curve should be regarded as a upper bound for the hole current in an (In,Ga)N MQW structure. This is in contrast to uni-polar electron transport, where alloy fluctuations and quantum corrections give rise to an \emph{increase} in the current when compared to a VCA result.~\cite{MiOD2021_JAP} Overall, we conclude that alloy disorder has a \textit{detrimental} effect
on \emph{hole} transport (In,Ga)N MQWs. The degree to which this impacts the I-V curve requires further careful research
into the description of the confining energy landscape.

\section{Conclusions}\label{sec:Conclusions}

In this work we applied the previously established TB-to-continuum framework to perform drift-diffusion calculations
for $p$-$i$-$p$ systems. The impact of alloy fluctuations was determined by comparing to a VCA, and quantum corrections
were included via LLT. Our results showed that alloy fluctuations have a detrimental effect on hole transport through In$_{0.1}$Ga$_{0.9}$N/GaN QW 
systems, although the degree to which this impacts results depends on the treatment of the localization landscape, and the smoothing applied.
For low Gaussian broadening values, $\sigma$, the alloy fluctuations reduce the current, due to the increased hole density localizing within the QWs and the resulting depletion of the barriers; this reduces
the conductivity in the barrier regions. When the landscape is heavily smoothed (large $\sigma$) this effect is reduced, and the I-V curve
approaches that of a smooth landscape (VCA). As already highlighted above, further studies on how to describe the (disordered) energy landscape are now required to shed more light onto the carrier transport in (In,Ga)N/GaN QW systems.


\section*{Acknowledgements}
The authors thank J. Fuhrmann (WIAS) for fruitful discussions. This work received funding from the Sustainable Energy Authority of Ireland and the Science Foundation Ireland (Nos. 17/CDA/4789 and 12/RC/2276 P2) and the Deutsche Forschungsgemeinschaft (DFG) under Germany’s Excellence Strategy EXC2046: MATH+, project AA2-15, as well as the Leibniz competition 2020.

\appendix
\section{Effective confining potential in MQW structure: Partitioned vs Unpartitioned LLT}\label{Appendix:LLT_partitioning_landscapes}
\begin{figure}[ht]
    \centering
    \includegraphics[width=0.5\textwidth]{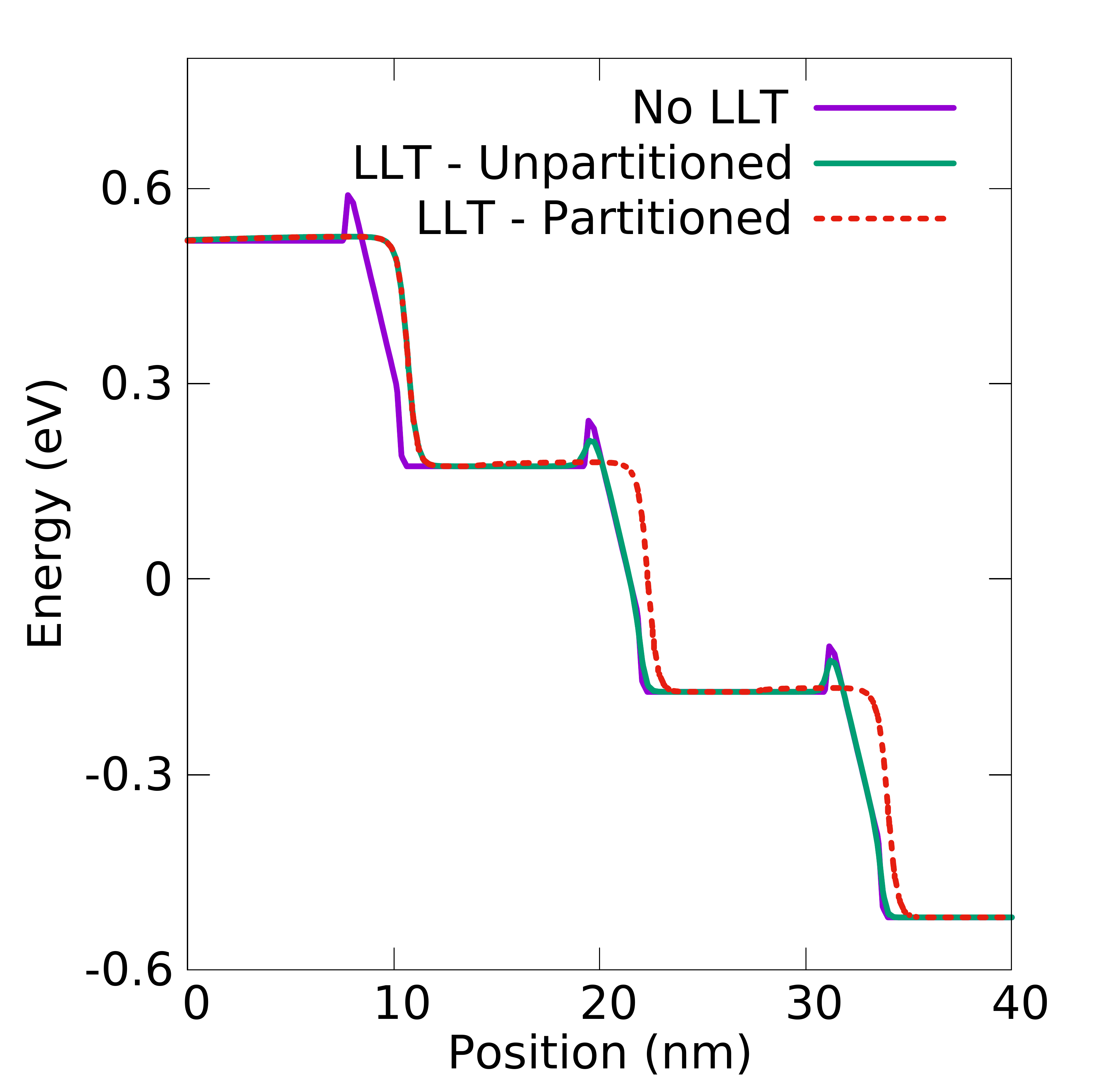}
    \caption{Valence band edge profile for a fictive (In,Ga)N/GaN multi-quantum well system in the absence of LLT (purple) and presence of LLT quantum corrections. When including LLT two scenarios are considered (i) using a single
    reference energy (green) and (ii) separate reference energies for each quantum well region (red, dashed).}
    \label{fig:MQWLLTcomparison}
\end{figure}

In this appendix we provide further insight into the question how the effective confining potential, $W$, obtained from LLT is modified when partitioning the MQW into sub-regions, i.e. different ``localization'' regions. As discussed in Sec.~\ref{subsec:problems_with_LLT_in_MQW}, the choice of the reference energy $E_\text{ref}$ for a given localization region can impact the resulting quantum corrected effective landscape.
As a test case we have the system discussed in the main part of the manuscript using a large potential
difference between the QWs (as shown in Fig.~\ref{fig:cartoon_psi}) forming the MQW. For demonstrative purposes we neglect any effects due to the presence of a $p$-$i$-$p$ junction
and we assume a capacitor-like potential profile with a potential drop across each QW of 0.35 V. 
Figure~\ref{fig:MQWLLTcomparison} reveals the impact that partitioning the MQW into different subregions has on the effective band edge. 
The starting point is the ``standard'' VCA description of the system VCA without quantum corrections (purple). Here, each QW exhibits the same VBE profile. Treating the MQW system as a single localization region within LLT, the resulting band edge profile (green) reveals that the band edge of
the first QW (leftmost in Fig.~\ref{fig:MQWLLTcomparison}) is smoothed significantly. However, the two other wells forming the MQW system, which are energetically far from the global reference energy chosen, undergo
noticeably smaller corrections. As discussed in the main text, this stems from the fact that the contributions from states in these QWs contribute only weakly to the series expansion of $u$
(Eqs.~(\ref{eq:u_expansion}) and (\ref{eq:expansion_coeffs})). However, Fig.~\ref{fig:MQWLLTcomparison} also reveals that when the system is partitioned into 3 sub-regions, and each localization region 
(QW region) is described by an individual reference energy which is close the \emph{local} ground state energy, the resulting effective landscape 
(red, dashed) is significantly softened in all three wells of the MQW system. In doing so, one assures that quantum corrections in all 3 QWs are properly treated. Figure~\ref{fig:MQWLLTcomparison} also shows that the landscape is not only smoothed but also continuous between each localization region, which is important to construct a global effective landscape that can be used for transport calculations.

\bibliography{lit.bib}
\bibliographystyle{apsrev}

\end{document}